\documentclass{article}
\PassOptionsToPackage{numbers, compress}{natbib}

\usepackage[preprint]{neurips_2026}

\makeatletter
\@ifpackageloaded{lineno}{\nolinenumbers}{}
\makeatother

\usepackage[utf8]{inputenc}
\usepackage[T1]{fontenc}
\usepackage{hyperref}
\usepackage{url}
\usepackage{booktabs}
\usepackage{amsfonts}
\usepackage{amsmath}
\usepackage{nicefrac}
\usepackage{microtype}
\usepackage[table,x11names]{xcolor}

\usepackage[pdftex]{graphicx}
\usepackage{siunitx}

\title{SCRAMJet: A Ultra Low Latency JetClass-Scale Transformer for the HL-LHC Level-1 Trigger}

\author{ 
  \bfseries 
  Aadeesh Sharma$^{1}$, 
  Chang Sun$^{2}$, 
  Lauri Laatu$^{1}$, 
  Arianna Cox$^{1}$, 
  Benedikt Maier$^{1}$, \\
  \bfseries
  Alexander Tapper$^{1}$, 
  Maria Spiropulu$^{2}$, 
  Wayne Luk$^{1}$, 
  Zhiqiang Que$^{3, 1}$\thanks{Corresponding author and and project advisor: z.que@bristol.ac.uk.}
  \\[1.0em]
  \begin{tabular}{c}
    $^{1}$Imperial College London, United Kingdom \\
    $^{2}$California Institute of Technology, USA \\
    $^{3}$University of Bristol, United Kingdom
  \end{tabular}
}

\begin{document}

\maketitle

\begin{abstract}
  Level-1 Triggers at the LHC classify collisions within a few microseconds on FPGAs, where a network is a fully unrolled circuit whose cost is area rather than time. We present SCRAMJet, a transformer encoder that replaces multi-head attention with a low-rank projection, trains under per-parameter learned fixed-point quantization, and compiles to Verilog with distributed arithmetic optimization. On the ten-class JetClass dataset it reaches $71.4 \pm 0.6\%$ accuracy with 0.157M multiply--accumulates, 6.1 accuracy points above the strongest published resource-constrained model at an eighth of the operations, and meets timing on a Virtex UltraScale+ VU13P at \SI{147}{\nano\second} using 25.7\% of the logic. We also show that resource-aware training silently removes attention channels: one whose learned bit-width reaches zero still carries signal during training but is absent from the compiled circuit,   so the loss never sees it go. We call these virtual channels, and a floor on the attention datalane that prevents them is worth 3.1 accuracy points at matched budget. Code and trained checkpoints are available at \url{https://github.com/Aaddy-1/hgq-jetformer}.
\end{abstract}

\section{Introduction}

Detectors at the CERN's Large Hadron Colliders (LHC) observe collisions at \SI{40}{\mega\hertz}, producing data at a rate
far beyond what can be recorded. A two-stage trigger reduces that rate, and its first
stage, the Level-1 Trigger (L1T), is implemented on FPGAs with an end-to-end latency
budget of a few microseconds \citep{cms_l1t_tdr}. An event the L1T does not select is
discarded permanently and cannot be recovered by any later analysis. Within that budget a
model must execute as a fully unrolled, spatially mapped circuit. To sustain a new
input every clock cycle within that budget, we compile the network as a fully unrolled,
spatially mapped circuit. There is no weight reuse and no sequential loop over layers,
since every multiplication in the network exists simultaneously in silicon. Model cost and deployment therefore must consider both physical space on the chip as well as latency,
and a design that is merely small in parameter count may still be unimplementable if its
arithmetic does not fit the device.

Transformers are difficult to use here. Offline they are the strongest models available,
with the Particle Transformer reaching 86.1\% accuracy on the JetClass dataset
\citep{particle_transformer}, but they require hundreds of millions of operations per jet,
three orders of magnitude beyond the trigger budget. Two recent
lines of work bring the problem within reach, and this work sits at their intersection.
\citet{submicro_transformers} demonstrate sub-microsecond transformer inference on FPGAs
by combining high-granularity quantization with distributed-arithmetic compilation,
reaching $\mathcal{O}(100)$\,\si{\nano\second} latency and 79.8\% accuracy on the five-class, three-feature
hls4ml benchmark. \citet{jetformer} introduce JetFormer, an encoder-only transformer
reaching 82.9\% on the ten-class JetClass dataset in floating point, but deploy only a compressed
eight-particle variant at \SI{2.7}{\milli\second} latency. We bridge these two lines by proposing
\textbf{SCRAMJet}, a Sub-microsecond Compiled Rank-reduced
Attention Model for Jet tagging. A budget enforced from the beginning also has to be allocated across the network during training. \citet{submicro_transformers} report attention channels collapsing under such training. This paper identifies the mechanism behind it and what it costs. i

To the best of our knowledge, this is the first demonstration of a quantized Transformer for JetClass that achieves sub-microsecond FPGA inference.
Our contributions are:

\begin{itemize}\setlength{\itemsep}{0pt}\setlength{\parskip}{0pt}
  \item \textbf{SCRAMJet}, a quantization-aware transformer for the JetClass dataset that
        compiles to
        bit-exact RTL and implements on a trigger-class FPGA inside the L1T budget, 6.1
        accuracy points above the strongest published model at an eighth of the operations.
  \item \textbf{Virtual channels}, a failure mode in resource-aware quantization-aware
        training. Channels whose learned precision reaches zero leave the compiled circuit but
        stay active during training. This accounts for the entire train--inference gap, and a floor on the
        attention datalane is worth 3.10 accuracy points at matched budget.
  \item \textbf{Two design rules that follow.} The resource surrogate can rank two designs
        in the opposite order to their compiled cost, so cost comes out by cutting channels
        rather than by lowering the budget.
\end{itemize}

\section{SCRAMJet}

\paragraph{Architecture.}
On a fully unrolled circuit every operation exists simultaneously in silicon, so cost is paid for structure that must physically exist rather than for operation count. SCRAMJet removes the two structures of JetFormer's encoder whose cost is not arithmetic: the $n \times n$ attention score matrix, which is paid for in area, and the class token, which is paid for in latency. A low-rank projection~\citep{linformer} replaces the score matrix with two $n \times k$ projections, linear in $n$. We swept the rank: at $k{=}1$ the softmax is over a single element and therefore identically one, reducing the block to a permutation-invariant set network with a global-context broadcast, with no loss to accuracy, but at the cost of 0.012 macro AUC, which is why we use a $k{=}2$ rank projection, so attention earns its
cost here in ranking quality rather than accuracy. Global average pooling replaces the class token: it aggregates after the blocks
instead of prepending a concatenation ahead of them, and the two are indistinguishable in accuracy. The proposed configuration projects
$n{\times}17$ constituent inputs to width $D{=}16$, passes them through a single block (2-head rank-2 attention and a $D{\to}2D{\to}D$
ReLU feed-forward network with residual adds and no normalization in the quantized path), and pools into a linear head
($D{\to}D{\to}10$).

\paragraph{Quantization and compilation.}
Every weight and activation carries its own learned fixed-point representation under
fine-grained mixed-precision quantization~\citep{hgq}. Bit-widths are trained jointly with the task
against a surrogate hardware cost, Effective Bit Operations(EBOPs)~\citep{hgq}, held at a setpoint by a controller. The trained model is then compiled to Verilog by the Alkaid~\cite{alkaid} compiler
together with distributed arithmetic (DA) optimization~\citep{da4ml},
which rewrites each constant-matrix--vector multiplication as a shared graph of additions
and shifts.

\section{Quantization Aware Training collapses the Attention block}
\label{sec:collapse}

\begin{figure}[htbp]
  \centering
  \includegraphics[width=0.82\linewidth]{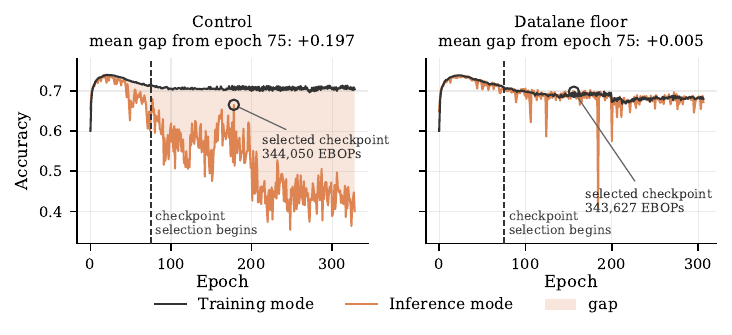}
  \caption{Training-mode and inference-mode accuracy for a matched-budget pair. The two
    modes compute different functions once a channel reaches zero width.}
  \label{fig:traingap}
\end{figure}

Training under an EBOPs budget produces a curve that appears healthy and a deployed
model that is markedly worse: accuracy plateaus near 72.3\% during training while the same
checkpoint in inference mode sits near 65.2\% (Figure~\ref{fig:traingap}). Averaged over every epoch after the controller
engages the separation is 19.7\%. This is not overfitting: one checkpoint in inference
mode on train, validation and test at 200{,}000 samples each spans 0.03 percentage points.

The training and inference forward passes compute different functions. During training the
quantizer applies rounding and returns; at inference it additionally applies a mask that
zeroes any channel whose total bit-width $k+i+f$ (sign, integer, and fractional bits) has fallen to zero or below. Masking these
channels is correct, since the hardware could not represent them in any case. The
difficulty is that during training those same channels carry signal normally. We call
these virtual channels: they contribute to training but require no hardware. The
optimizer is therefore free to compress a channel until it is virtual without the task
loss ever registering a cost, and the training metric goes on describing virtual channels
absent from the generated circuit.

Disabling the mask at inference closes the gap exactly: the same checkpoint scores
65.2\% with the mask on and 72.35\% with it off, against a training-mode metric of
72.32\%, agreeing to 0.03 percentage points, so virtual channels account for all of it. It is also
localised. At the 500{,}000-EBOPs budget, 7{,}015 of 80{,}868 datalane quantizer
parameters sit at or below zero width, every one inside attention, with 60.0\% of the
Linformer key projection's input channels virtual against 0.0\% elsewhere.

\section{Protecting the attention block, and reducing the compiled cost}

\paragraph{Protect the attention datalane with a bit-width floor.}
The collapse has been reported before \citep{submicro_transformers}, in a setting that
already had a bit-width constraint meant to prevent it, and we inherited that constraint
from them. It bounds a parameter existing only in the representation used for \emph{weight} quantizers, while the
quantizers at risk here are \emph{activation} quantizers in a different one. A floor on
the parameters that do govern them removes the deletion. Holding seed, commit and budget
fixed and varying only the floor, the two arms land within 0.12\% on EBOPs (344{,}050
against 343{,}627), accuracy rises from 66.38\% to 69.48\%, and the
virtual-channel count falls from 7{,}800 to zero without the pruning moving elsewhere. It
also makes accuracy and compiled cost roughly four and twelve times more reproducible.

\begin{figure}[htbp]
  \centering
  \includegraphics[width=0.82\linewidth]{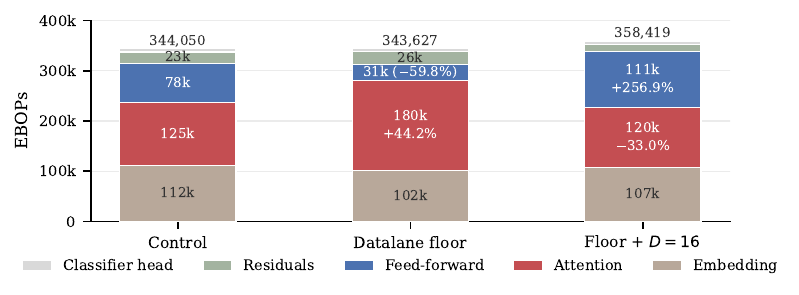}
  \caption{Per-section EBOPs. The floor raises the attention budget by 44.2\%, paid for
    by a 59.8\% cut to the feed-forward network; halving the width moves it back.}
  \label{fig:budgetalloc}
\end{figure}

The budget is fixed, so protecting attention means recovering it elsewhere. Figure~\ref{fig:budgetalloc} measures the per-layer allocation before and
after: attention gains 44.2\% of its EBOPs, from 124{,}527 to 179{,}525, and the
controller finds that budget by cutting the feed-forward network by 59.8\% and the
embedding by 9.2\%. None of this reallocation was specified by hand.

\paragraph{Reduce the model width.}
The same matched pair shows that EBOPs is an unreliable guide to hardware cost. Matched
to within 0.12\% on EBOPs, the control compiles to 784{,}360 adder-graph units and the
floored design to 1{,}114{,}938, 42.1\% more, even though the floored design is the
cheaper of the two by EBOPs. EBOPs sums products
of operand widths, so narrowing a channel makes every node slightly cheaper, while
deleting one removes its whole fan-out subtree from the netlist. EBOPs cannot tell a
channel narrowed to nothing apart from one that is absent. Lowering the setpoint is therefore not a reliable reduction in compiled cost;
removing channels is --- the same removal that makes a channel virtual, but chosen by
the designer rather than left to the optimizer. Halving the width from $D{=}32$ to $D{=}16$ cost no accuracy, raised EBOPs
by 4.3\%, and removed 38.8\% of the compiled cost.

\paragraph{Crop the constituent axis from 128 to 64.}
The JetClass dataset pads every jet to 128 constituents and an unrolled circuit pays for
padding in
full. Most of it is empty: constituents are stored in order of decreasing momentum and the
median jet has 38, so cropping at 64 leaves 93.7\% of jets untouched and retains 99.887\%
of total jet $p_T$. The freed budget goes to precision in the channels that remain.

\section{Results}

Models are trained on a two-million-jet subset of the JetClass dataset
\citep{jetclass_dataset} and evaluated on two million held-out jets, following the resource-constrained benchmark protocol and
baselines of \citet{phatjet} (note \citet{submicro_transformers} and \citet{jetformer} report no trigger-scale JetClass models).
Because unfloored QAT is seed-sensitive, we report mean $\pm$ standard deviation over $n=3$ fixed seeds.

\begin{table}[htbp]
  \centering
  \caption{Resource-constrained models on the JetClass dataset. Prior-work figures are single
    published
    values from \citet{phatjet} on the same training subset. MACs are
    multiply--accumulates.}
  \label{tab:comparison}
  \footnotesize
  \setlength{\tabcolsep}{5pt}
  \begin{tabular}{lcccrr}
    \toprule
    Model                          & Const.        & Accuracy (\%)             & Macro AUC           & Params & MACs  \\
    \midrule
    PHAT-JeT \citep{phatjet}       & 128           & 65.38                     & 0.9391              & 6.8K   & 1.24M \\
    Linformer baseline             & 128           & 62.64                     & 0.9305              & 9.5K   & 1.49M \\
    SAL-T \citep{salt}             & 128           & 62.19                     & 0.9291              & 7.0K   & 1.42M \\
    JEDI-linear \citep{jedilinear} & 128           & 59.52                     & 0.9192              & 6.1K   & 1.31M \\
    \midrule
    SCRAMJet (128 const.)          & 128           & $68.61 \pm 2.17$          & $0.9269 \pm 0.0013$
                                   & 3.4K          & 0.314M                                                           \\
    \textbf{SCRAMJet}              & 64            & $\mathbf{71.43 \pm 0.62}$ & $0.9289 \pm 0.0037$
                                   & \textbf{3.1K} & \textbf{0.157M}                                                  \\
    \bottomrule
  \end{tabular}
\end{table}

\paragraph{Accuracy.}
SCRAMJet is 6.1 accuracy points above the strongest published model and 8.8 above their
Linformer baseline, the direct architectural analogue, at an eighth of the operation
count. The margin does not depend on the shorter constituent axis: matched at 128
constituents it still leads by 3.2 points at a quarter of the operations. Cropping
improves both metrics. On macro AUC it sits 0.0102 below the best prior model, so it leads on
accuracy and trails on ranking quality. One difference is
not in
Table~\ref{tab:comparison}: every prior figure is floating-point simulation, while
SCRAMJet's is a fixed-point design verified against the generated RTL and routed.

\begin{table}[htbp]
  \centering
  \caption{Implementation results on \texttt{xcvu13p-flga2577-2-e}. Both designs are
    $D{=}16$ with the datalane floor and differ only in the constituent axis, and
    both are fully routed and meet all timing constraints.}
  \label{tab:pnr}
  \footnotesize
  \begin{tabular}{lrr}
    \toprule
    Resource            & 64 constituents                                           & 128 constituents    \\
    \midrule
    CLB LUTs            & 443{,}904 (25.69\%)                                       & 532{,}275 (30.80\%) \\
    CLB registers       & 187{,}774 (5.43\%)                                        & 232{,}498 (6.73\%)  \\
    DSP blocks          & 8 (0.07\%)                                                & 3 (0.02\%)          \\
    Block RAM           & 0 (0.00\%)                                                & 0 (0.00\%)          \\
    Latency             & 21 cy.\ @ \SI{7.0}{\nano\second} = \SI{147}{\nano\second}
                        & 20 cy.\ @ \SI{7.0}{\nano\second} = \SI{140}{\nano\second}                       \\
    Total on-chip power & \SI{32.0}{\watt}                                          & \SI{40.8}{\watt}    \\
    \bottomrule
  \end{tabular}
\end{table}

\paragraph{Hardware.}
The generated Verilog reproduces the trained model exactly: zero mismatches over
1{,}310{,}720 compared outputs. The design has no block RAM and no
design-rule violations, and runs in 21 cycles
at \SI{7.0}{\nano\second} for a latency of \SI{147}{\nano\second}: more than an order of
magnitude inside the L1T budget, and the same
$\mathcal{O}(100)$\,\si{\nano\second} range as the only comparable deployed transformer
\citep{submicro_transformers}, whose \SI{78}{\nano\second} is at 64 constituents on the
five-class hls4ml benchmark. Against
the 128-constituent design the crop removes 16.6\% of the logic and 21.6\% of the power
while gaining 2.8 accuracy points.

\section{Conclusion}
SCRAMJet reaches 71.4\% on the ten-class JetClass dataset, compiles to bit-exact RTL, and
implements
on a trigger-class FPGA at \SI{147}{\nano\second} using a quarter of the device. A
virtual channel costs the compiler almost nothing and the surrogate cannot see the
difference, which is a property of resource-aware training in general and of the other L1T tasks under the same
budget.



{\bf Acknowledgment.}  
Work done by Imperial College is funded by the Science and Technology Facilities Council (STFC) grant ST/W000636/1 and by UKRI grant UKRI256.
Work done at Caltech is partially supported by by United States DoE (grant numbers DE-SC0011925, DE-FOA-0002705) and NSF (grant numbers PHY240298, PHY2117997).

\bibliographystyle{unsrtnat}
\bibliography{scramjet_refs}

\end{document}